\documentclass[a4paper,fleqn,usenatbib]{mnras}

\usepackage{newtxtext,newtxmath}

\usepackage[T1]{fontenc}
\usepackage{ae,aecompl}
\usepackage{enumitem}

\usepackage{graphicx}	
\usepackage{amsmath}	
\usepackage{amssymb}	

\defcitealias{Peng:2010aa}{P10}
\defcitealias{Peng:2012aa}{P12}
\defcitealias{Treister:2012ab}{T12}
\defcitealias{Ajello:2012aa}{A12}
\defcitealias{Weigel:2017aa}{W17a}
\defcitealias{Weigel:2017ab}{W17b}

\newcommand{\lambdastarX}{$-1.84^{+0.30}_{-0.37}$}
\newcommand{\lambdastarXnoerror}{$-1.84$}
\newcommand{\deltaXa}{$0.47^{+0.20}_{-0.42}$}
\newcommand{\deltaXanoerror}{$0.47$}

\newcommand{\deltaXb}{$2.53^{+0.68}_{-0.38}$}
\newcommand{\deltaXbnoerror}{$2.53$}
\newcommand{\xistarX}{$-1.65$}

\newcommand{\modelAredchifrac}{$26.8$}
\newcommand{\modelAredchixlf}{$0.3$}
\newcommand{\modelBredchifrac}{$14.3$}
\newcommand{\modelBredchixlf}{$3.0$}
\newcommand{\modelCredchifrac}{$16.3$}
\newcommand{\modelCredchixlf}{$0.6$}
\newcommand{\modelDredchifrac}{$2.5$}
\newcommand{\modelDredchixlf}{$0.4$}
\newcommand{\modelEredchifrac}{$1.8$}
\newcommand{\modelEredchixlf}{$4.9$}
\newcommand{\modelFredchifrac}{$2.1$}
\newcommand{\modelFredchixlf}{$12.7$}

\newcommand{\modelAxistarratio}{$1.00$}
\newcommand{\modelBxistarratio}{$1.93$}
\newcommand{\modelCxistarratio}{$0.33$}
\newcommand{\modelDxistarratio}{$10.00$}
\newcommand{\modelExistarratio}{$3.31$}
\newcommand{\modelFxistarratio}{$19.32$}

\title[The AGN merger fraction]{The fraction of AGN in major merger galaxies and its luminosity dependence}

\author[Anna K. Weigel et al.]
	{\parbox[t]{18cm}{
	Anna K. Weigel,$^{1}$\thanks{E-mail: anna.weigel@phys.ethz.ch}
	Kevin Schawinski,$^{1}$
	Ezequiel Treister,$^{2}$
	Benny Trakhtenbrot$^{3}$\thanks{Zwicky Fellow.} and 
	David B. Sanders$^{4}$
	\\}
	\\
	$^{1}$Institute for Particle Physics and Astrophysics, ETH Zurich, Wolfgang-Pauli-Strasse 27, CH-8093 Zurich, Switzerland\\
	$^{2}$Instituto de Astrof\'isica, Facultad de F\'isica, Pontificia Universidad Cat\'olica de Chile, 306, Santiago 22, Chile\\
	$^{3}$Department of Physics, ETH Zurich, Wolfgang-Pauli-Strasse 27, CH-8093 Zurich, Switzerland\\
	$^{4}$Institute for Astronomy, 2680 Woodlawn Drive, University of Hawaii, Honolulu, HI 96822, USA\\
}

\date{Accepted XXX. Received YYY; in original form ZZZ}

\pubyear{2017}

\begin{document}
\label{firstpage}
\pagerange{\pageref{firstpage}--\pageref{lastpage}}
\maketitle

\begin{abstract}
	We use a phenomenological model which connects the galaxy and AGN populations to investigate the process of AGN triggering through major galaxy mergers at $z\sim0$. The model uses stellar mass functions as input and allows the prediction of AGN luminosity functions based on assumed Eddington ratio distribution functions (ERDFs). We show that the number of AGN hosted by merger galaxies relative to the total number of AGN increases as a function of AGN  luminosity. This is due to more massive galaxies being more likely to undergo a merger and does not require the assumption that mergers lead to higher Eddington ratios than secular processes. Our qualitative analysis also shows that to match the observations, the probability of a merger galaxy hosting an AGN and  accreting at a given Eddington value has to be increased by a factor $\sim10$ relative to the general AGN population. An additional significant increase of the fraction of high Eddington ratio AGN among merger host galaxies leads to inconsistency with the observed X-ray luminosity function. Physically our results imply that, compared to the general galaxy population, the AGN fraction among merger galaxies is $\sim 10$ times higher. On average, merger triggering does however not lead to significantly higher Eddington ratios.
\end{abstract}




\section{Introduction}
Active galactic nuclei (AGN) are often thought be an essential part of the classical galaxy major merger sequence \citep{Sanders:1988aa}. Due to the merger, gravitational torques drive gas to the center of the newly formed galaxy \citep{Mihos:1996aa}. Radiative and/or mechanical output from the accreting super massive black hole \citep{Fabian:2012aa}, possibly in combination with feedback from star formation, leads to the heating and expulsion of this gas, quenching star formation and allowing the now elliptical galaxy to rapidly transition to the red sequence \citep{Di-Matteo:2005aa, Springel:2005ab, Kaviraj:2011aa, Schawinski:2014aa, Smethurst:2015aa}. 

The AGN merger fraction, that is the number of AGN found in major merger galaxies relative to the total number of AGN, allows us to investigate the nature of the AGN - major merger relation. For luminous quasars, evidence for \citep{Sanders:1988aa, Bahcall:1997aa, Veilleux:2009aa, Hong:2015aa} and against  \citep{Schawinski:2012aa} merger triggering has been found. At more moderate AGN luminosities, signs of recent major galaxy mergers have been found to be less frequent or lacking \citep{De-Robertis:1998aa, Bennert:2011aa, Kocevski:2012aa}, with the AGN fraction among merger and non-merger galaxies being similar \citep{Gabor:2009aa, Cisternas:2011aa}. By combining AGN merger fraction measurements over the $0 < z < 3$ range, \citeauthor{Treister:2012ab} (\citeyear{Treister:2012ab}, hereafter \citetalias{Treister:2012ab}) showed that the fraction of AGN in major merger galaxies increases from $\sim3\%$ to $\sim100\%$ within the $10^{43}- 10^{47}\ \rm erg\ s^{-1}$ bolometric luminosity range. \cite{Glikman:2015aa} show that this trend also holds for $z\sim2$ dust-reddened quasars at the most extreme AGN luminosities (also see e.g \citealt{Banerji:2017aa}). This seems to suggest that galaxy mergers may be driving the episodes of fastest super massive black hole growth. We note that many studies, including the aforementioned ones, naturally suffer from limitations related to survey design, observational biases, methods used for merger identification, and/or  control samples (or lack thereof). These may have contributed, to some extend, to the confusion regarding the links between major mergers and AGN.

The phenomenological model by \citeauthor{Weigel:2017aa} (\citeyear{Weigel:2017aa}, hereafter \citetalias{Weigel:2017aa}) describes the relation between the galaxy population, namely the stellar mass function, and the AGN population, namely the AGN luminosity function. The two populations are connected through the seemingly universal (also see e.g. \citealt{Aird:2012aa}), mass independent Eddington ratio distribution function (ERDF, $\xi(\lambda_{\rm Edd})$, $\lambda_{\rm Edd} = L_{\rm bol}/L_{\rm Edd}$). We use this model to examine three aspects of the AGN - major merger connection. First, we investigate the luminosity dependence of the AGN merger fraction. Second, we consider the difference between the fraction of major merger galaxies that host AGN and the fraction of AGN found among the general galaxy population. Third, we explore possible shapes for the ERDF of AGN in major merger galaxies. Our analysis is based on a qualitative approach. A sophisticated statistical analysis is not justified given current observational uncertainties and the number of assumptions that our approach is based on. Nonetheless, we can exclude extreme scenarios and show what must broadly be true for AGN in major merger galaxies. Throughout this work we assume a $\Lambda$CDM cosmology with $h_0 = 0.7$, $\Omega_m = 0.3$ and $\Omega_\Lambda = 0.7$.\\
\section{Analysis}
\subsection{The ERDF model and the major merger mass function}
Our analysis is based on the galaxy major merger mass function by \cite{Weigel:2017ab} (hereafter \citetalias{Weigel:2017ab}) and the ERDF model by \citetalias{Weigel:2017aa}. \citetalias{Weigel:2017ab} use a sample of visually selected major merger systems (mass ratio up to 1:3) from Galaxy Zoo 1 \citep{Darg:2010aa, Lintott:2008aa,Lintott:2011aa} to determine the stellar mass function of local ($0.02 < z < 0.06$) major merger galaxies.  The mass function is determined by using the method outlined by \cite{Weigel:2016aa} and is well described by a single Schechter function:
\begin{equation}
\Phi(M) d\log M = \ln(10) \Phi^{*} e^{-M/M^{*}} \left(\frac{M}{M^{*}}\right)^{\alpha+1} d\log M.
\end{equation}
For the stellar mass function of major merger galaxies, $\Phi_{\rm m}(M)$, \citetalias{Weigel:2017ab} report the following best-fitting values $\log (M^{*}/M_\odot)$ = $10.89\pm0.06$, $\log (\Phi^{*}/\rm Mpc^{-3})$ = $-4.76\pm0.03$, and  $\alpha$ = $-0.55\pm0.08$.

The \citetalias{Weigel:2017aa} model uses a forward modeling approach to predict the \textit{AGN} luminosity function from the \textit{galaxy} stellar mass function. \citetalias{Weigel:2017aa} show that the properties of the local ($z\lesssim 0.1$) AGN population can be described by two independent ERDFs: one for X-ray and one for radio detected AGN. The two ERDFs differ in shape, however they are both mass independent (see e.g. \citealt{Jones:2017aa, Bernhard:2018aa} for higher $z$). This implies a mass independent intrinsic AGN fraction (also see e.g. \citealt{Aird:2012aa}). Compared to lower mass galaxies, more massive galaxies are neither more likely to host AGN nor to host high accretion rate AGN. 

\citetalias{Weigel:2017aa} use a constant black hole mass to stellar mass ratio ($\log (M_{\rm BH}/M) = -2.75$, scatter 0.3 dex, e.g.  \citealt{Haring:2004aa, Jahnke:2009aa, Kormendy:2013ab, McConnell:2013aa, Marleau:2013aa, Reines:2015aa}) and a constant bolometric correction (for hard X-rays: $\log L_{\rm bol} = 10\times L_{\rm{15 - 55 \rm keV}}$, \citealt{Rigby:2009aa, Vasudevan:2009aa}). They assume that X-ray selected AGN are primarily found in optically blue and green galaxies (e.g. \citealt{Hickox:2009aa, Schawinski:2009ab, Treister:2009ab, Koss:2011aa, Rosario:2013ab, Goulding:2014aa}) and use the corresponding stellar mass function as input for the model. The best-fitting double Schechter function is given by 
\begin{equation}
\begin{aligned}
\Phi(M) d\log M =& \ln(10) e^{-M/M^{*}} \\
										   & \times \left[\Phi^{*}_{1} \left(\frac{M}{M^{*}}\right)^{\alpha_{1} + 1} + \Phi^{*}_{2} \left(\frac{M}{M^{*}}\right)^{\alpha_{2} + 1} \right]\\ & d\log M.
\end{aligned}
\end{equation}
For $\Phi_{\rm bg}(M)$, the mass function of optically blue and green galaxies, \citetalias{Weigel:2017aa} report the following best-fitting parameters $\log (M^{*}/M_\odot)$ = $10.67\pm0.02$, $\log (\Phi^{*}_{1}/\rm Mpc^{-3})$ = $-3.10\pm0.10$,  $\alpha_{1}$ = $-1.38\pm0.05$,  $\log (\Phi^{*}_{2}/\rm Mpc^{-3})$ = $-2.91\pm0.18$, and  $\alpha_{2}$ = $-0.70\pm0.11$.

 For the ERDF a broken powerlaw shape of the following form is assumed
\begin{equation}
\xi(\lambda) = \frac{dN}{Nd\log\lambda} = \xi^{*} \times \left[\left(\frac{\lambda}{\lambda^{*}}\right)^{\delta_1} + \left(\frac{\lambda}{\lambda^{*}}\right)^{\delta_2}  \right]^{-1}.
\end{equation}
Here, $\lambda^{*}$, $\delta_1$ and $\delta_2$ correspond to the characteristic break, and the low and the high Edddington ratio slopes, respectively. 

The prediction of the AGN luminosity function consists of two steps. First, the observed stellar mass function is convolved with a normal distribution to predict the black hole mass function. Second, the black hole mass function is convolved with the assumed Eddington ratio distribution function to determine the bolometric AGN luminosity function \citep{Caplar:2015aa}. With this simple, phenomenological model, \citetalias{Weigel:2017aa} illustrate, that the observed AGN hard X-ray luminosity function (XLF, \citealt{Ajello:2012aa}, hereafter \citetalias{Ajello:2012aa}) can be reproduced by a mass independent, broken power law shaped ERDF with the following parameters: $\log \lambda_{\rm Edd}^{*} =$ \lambdastarX and  $\delta_1 = $ \deltaXa, $\delta_2 =$ \deltaXb, $\log \xi^{*} =$ \xistarX\footnote{By following the \citetalias{Weigel:2017aa} method, the normalization of the best-fitting ERDF, $\xi^{*}$, depends on the assumed Eddington ratio range. \citetalias{Weigel:2017aa} use $\log \lambda_{\rm Edd, min} = -8$, $\log \lambda_{\rm Edd, max} = 1$, whereas we show Eddington ratios in the range $-5 < \log \lambda_{\rm Edd} < 1$. However, our results are independent of the absolute value which is assumed for $\xi^{*}$.}. A mild mass dependence of the ERDF is permitted by the observations. However, such a $\xi(\lambda, \log M)$ model has to comply with strict constraints, the introduction of a mass dependence requires further assumptions, and a mass independent ERDF is the simplest and most straightforward solution. The observations are still consistent with a mass independent ERDF, even if a luminosity dependent bolometric correction is introduced. \citetalias{Weigel:2017aa} also test and discuss the effects of lognormal or Schechter function shaped ERDFs (e.g. \citealt{Kollmeier:2006aa, Hopkins:2009aa, Conroy:2013aa, Trump:2015aa, Schulze:2015aa, Jones:2016aa, Bongiorno:2016aa}).  

\subsection{Method}
\begin{figure*}
	\includegraphics[width=\textwidth]{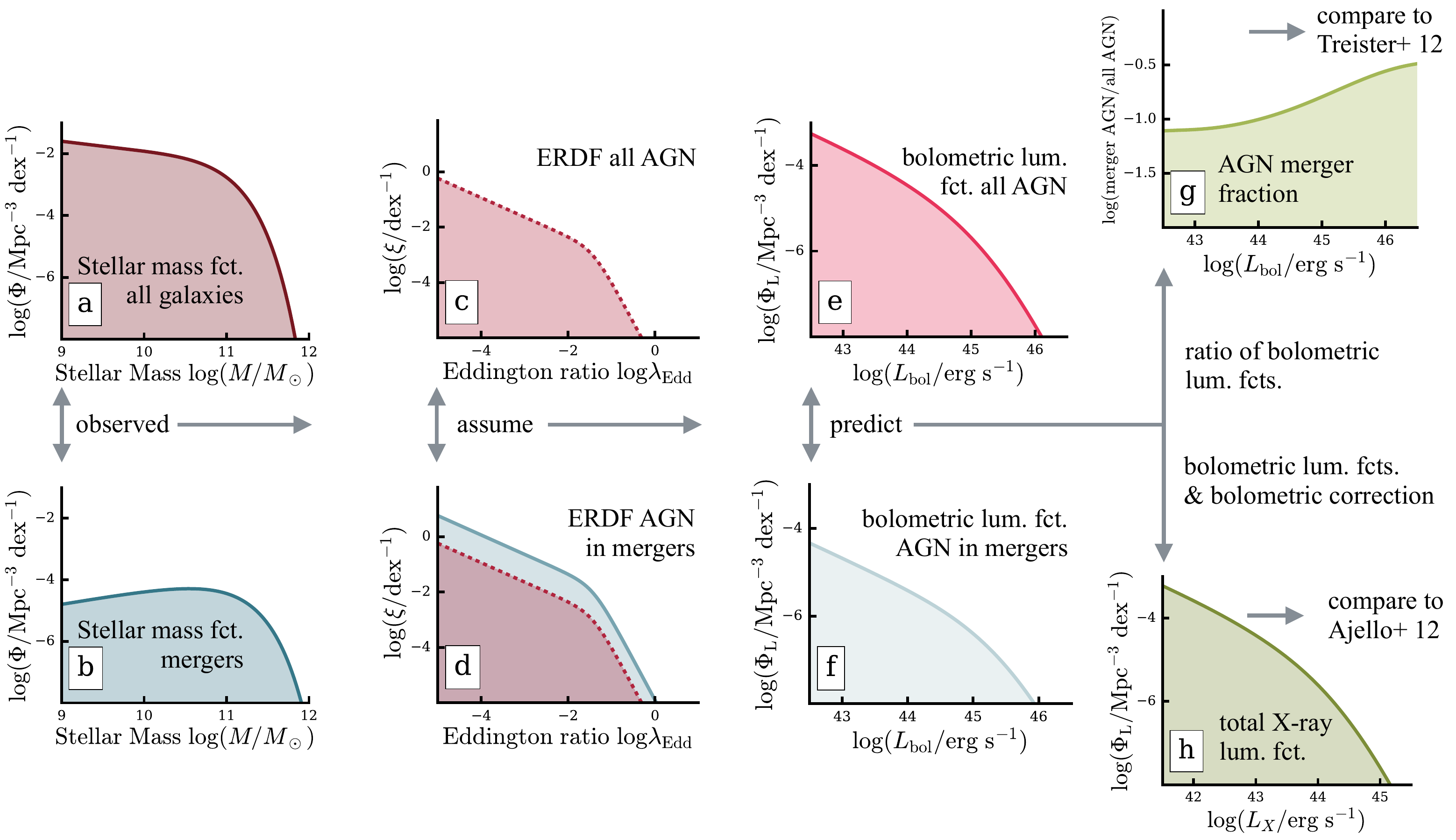}
	\caption{\label{fig:cartoon}Schematic illustration of our method. Following the approach by \citetalias{Weigel:2017aa}, we use an observed \textit{galaxy} stellar mass function and an assumed ERDF as input to predict the bolometric \textit{AGN} luminosity function. We use the stellar mass function of optically blue and green galaxies (panel a)  and the best-fitting X-ray ERDF by \citetalias{Weigel:2017aa} (panel c)  to derive the bolometric luminosity function of all radiatively efficient AGN ($\Phi_{\rm L, all}(L_{\rm bol})$, panel e). We use the stellar mass function of optically selected major merger galaxies (panel b, \citetalias{Weigel:2017ab}) and an assumed ERDF (panel d) to determine the bolometric luminosity function of AGN in major galaxy mergers ($\Phi_{\rm L, m}(L_{\rm bol})$, panel f). We compute the ratio of $\Phi_{\rm L, m}(L_{\rm bol})$ and $\Phi_{\rm L, all}(L_{\rm bol})$ to predict the AGN merger fraction as a function of bolometric luminosity (panel g).  We also derive the XLF of all AGN (panel h) by applying a bolometric correction and a rescaling factor. We compare our predictions to the results by \citetalias{Treister:2012ab} and \citetalias{Ajello:2012aa}.}
\end{figure*}

We combine the results by \citetalias{Weigel:2017aa} and \citetalias{Weigel:2017ab} to study the properties of AGN in major merger galaxies. We assume that AGN in major merger galaxies accrete via radiatively efficient accretion. We treat these `merger AGN' as a subgroup of all AGN that accrete via radiatively efficient accretion, which we refer to as the `the general AGN population' or as `all AGN'. We assume that all the AGN we are considering are detected in the hard X-rays (e.g.  \citealt{Mushotzky:2004aa, Brandt:2005aa}). 

We do not include radio selected AGN in our analysis. According to \cite{Ellison:2015ab}, local ($0.01  < z < 0.2$) radio AGN are likely to be fueled by large-scale environmental effects or galaxy internal processes (however see \citealt{Chiaberge:2015aa}  for $z > 1$ radio AGN).

For the general AGN population we use the ERDF of X-ray selected AGN which was determined by \citetalias{Weigel:2017aa}. For merger AGN we also assume a mass independent, broken powerlaw shaped ERDF.

We use the method by \citetalias{Weigel:2017aa}, illustrated in Fig. \ref{fig:cartoon}, to predict the bolometric AGN luminosity functions of all AGN and of merger AGN. The stellar mass function of optically blue and green galaxies (panel a) and the best-fitting X-ray ERDF by \citetalias{Weigel:2017aa} ($\xi_{\rm all}(\lambda_{\rm Edd})$, panel c) allow us to predict the bolometric luminosity function of all AGN ($\Phi_{\rm L, all}(L_{\rm bol})$, panel e). We use the major merger mass function by \citetalias{Weigel:2017ab} (panel b), combined with an assumed ERDF ($\xi_{\rm m}(\lambda_{\rm Edd})$, panel d) to obtain the bolometric luminosity function of merger AGN ($\Phi_{\rm L, m}(L_{\rm bol})$, panel f).

In addition to the bolometric AGN luminosity functions, we determine the AGN merger fraction as a function of bolometric luminosity (panel g). We define the AGN merger fraction as the space density  of AGN in major merger galaxies relative to the space density of all AGN. By using the bolometric luminosity functions, the AGN merger fraction can be expressed as: 
\begin{equation}\label{eq:merger_frac}
\rm AGN\ merger\ fraction (\log L) = \frac{\Phi_{\rm L, m}(\log L)}{\Phi_{\rm L, all}(\log L)}.
\end{equation}

For the stellar mass functions we are considering the $9 < \log(M/M_{\odot}) < 12$ range, which allows us to constrain the bolometric luminosity function in the $42.5 < \log(L_{\rm bol}/\rm erg\ s^{-1}) < 46.6$ and the XLF in the $41.5 < \log(L_{X}/\rm erg\ s^{-1}) < 45.6$ range (see discussion in Appendix B.1. of \citetalias{Weigel:2017aa}).  

We also estimate the total XLF of all merger and non-merger AGN (panel h). To account for the fact that merger AGN are treated as a subgroup of the general AGN population, we compute the bolometric luminosity function of non-merger AGN by rescaling $\Phi_{\rm L, all}(L_{\rm bol})$. For the rescaling we use the space densities of major mergers and blue and green galaxies given by their respective stellar mass functions. After applying the bolometric correction, the total XLF is given by the sum of $\Phi_{\rm L, non-m}$ and $\Phi_{\rm L, m}$. 

Motivated by \citetalias{Weigel:2017aa}, we use a \textit{constant} bolometric correction, $k_{\rm bol, X}$, to derive the total XLF. \citetalias{Weigel:2017aa} discuss the effect of a possible luminosity dependence of the bolometric correction on their results (see their Sect. B.2). The observed XLF \citepalias{Ajello:2012aa} is still consistent with a mass independent, broken power law shaped ERDF, even if the ratio between bolometric and hard X-ray luminosity is assumed to vary as a function of luminosity. Compared to the case where $k_{\rm bol, X}$ is constant, the slopes of the ERDF change, while $\lambda^{*}$ stays essentially constant. As we will discuss below, we consider $\xi_{\rm m}(\lambda)$ relative to $\xi_{\rm all}(\lambda)$. In the context of the present work, changing the slopes of $\xi_{\rm all}(\lambda)$ would thus imply the same relative change of $\delta_1$ and $\delta_2$ for $\xi_{\rm m}(\lambda)$, keeping the ratio between these two ERDFs and the bolometric luminosity functions constant. The AGN merger fraction, which depends on the ratio of $\Phi_{\rm L, m}(\log L)$ and $\Phi_{\rm L, all}(\log L)$ (see equation \ref{eq:merger_frac}), would thus be unaffected by a luminosity dependent bolometric correction. Similarly, our results for the total XLF would not be affected,  as $\Phi_{\rm L, m}$ and $\Phi_{\rm L, all}$ experience the same mapping between bolometric and hard X-ray luminosity and \citetalias{Weigel:2017aa} have shown that the observations are still consistent with a mass independent ERDF. We thus conclude that our present analysis and results are not affected by a a possible luminosity dependence of the X-ray bolometric correction.
 
 Besides the AGN merger fraction, our method also allows us to determine the `AGN host galaxy merger fraction'.  This quantity expresses the  AGN fraction among major merger galaxies relative to the AGN fraction among the general galaxy population. To compute the AGN fraction, we take the ratio of the space density of AGN, given by the integral over the AGN luminosity function, to the space density of galaxies, given by the integral over the galaxy stellar mass function. According to \citetalias{Weigel:2017aa}, the AGN fraction is also proportional to the product of  $\xi^{*}/\xi^{*}_{\rm norm}$ and the integral over the stellar mass function. Here $\xi^{*}_{\rm norm}$ is the normalization of the ERDF after it has been normalized to 1. We can thus express the AGN host galaxy merger fraction in the following way:
 \begin{equation}\label{eq:host_gal_merg_frac}
\begin{aligned}
 \rm AGN\ host\ &\rm galaxy\ merger\ fraction \\
 &= \frac{\int \Phi_{\rm L, m} d\log L/\int \Phi_{\rm m}d\log M}{\int \Phi_{\rm L, all} d\log L/\int \Phi_{\rm bg} d\log M}\\
 &= \frac{(\xi^{*}_{\rm m}/\xi^{*}_{\rm m, norm}) \times \int \Phi_{\rm m} d\log M/\int \Phi_{\rm m}d\log M}{(\xi^{*}_{\rm all}/\xi^{*}_{\rm all, norm}) \times \int \Phi_{\rm bg}d\log M/\int \Phi_{\rm bg} d\log M}\\
 &= \frac{\xi^{*}_{\rm m}/\xi^{*}_{\rm m, norm}}{\xi^{*}_{\rm all}/\xi^{*}_{\rm all, norm}}\\ 
 &=\frac{\xi^{*}_{\rm m}\times \int \xi_{\rm m}(\lambda_{\rm Edd}, \xi^{*} = 1) d\log \lambda_{\rm Edd}}{\xi^{*}_{\rm all} \times \int \xi_{\rm all}(\lambda_{\rm Edd}, \xi^{*} = 1) d\log \lambda_{\rm Edd}}\\
 &=\frac{\int \xi_{\rm m}(\lambda_{\rm Edd}) d\log \lambda_{\rm Edd}}{\int \xi_{\rm all}(\lambda_{\rm Edd}) d\log \lambda_{\rm Edd}}.
\end{aligned}
 \end{equation}
 Here, the integrals are computed over the considered $\log L$, $\log M$, and $\log \lambda_{\rm Edd}$ ranges.  Below, we will, for instance, increase the integral over the merger ERDF by a factor 10 relative to the integral over the ERDF for the general AGN population. According to equation \ref{eq:host_gal_merg_frac}, $\int \xi_{\rm m} (\lambda) d\log \lambda_{\rm Edd} = 10\times \int \xi_{\rm all}(\lambda_{\rm Edd}) d\log \lambda_{\rm Edd}$ implies that the fraction of AGN among major merger galaxies is ten times higher than among the general galaxy population. 

To estimate the statistical errors on the AGN luminosity functions and the AGN merger fraction, we take the errors on the input stellar mass functions into account. We assume that the errors on the best-fitting Schechter function parameters follow a normal distribution and randomly draw $\log M^{*}$, $\log \Phi^{*}$, and $\alpha$ values. For each representation of the input stellar mass functions we determine the AGN luminosity functions and the AGN merger fraction following the method outlined above. We then compute the mean $\log \Phi_{\rm L}$ and  $\log$(AGN merger fraction) values and corresponding standard deviations of all realizations. 

At high luminosities especially, the uncertainty on $\log$(AGN merger fraction) is significantly affected by the shape of the merger AGN ERDF. If $\xi_{\rm m}(\lambda_{\rm Edd})$ has a higher break $\lambda_{\rm Edd}^{*}$ or shallower slope $\delta_2$ than $\xi_{\rm all}(\lambda_{\rm Edd})$, the errors on the AGN fraction at high luminosities are smaller than in the case where $\xi_{\rm m}(\lambda_{\rm Edd})$ and $\xi_{\rm all}(\lambda_{\rm Edd})$ are similar in shape. With a higher $\lambda_{\rm Edd}^{*}$ or shallower $\delta_2$ value, the massive end of the merger mass functions and the corresponding uncertainties are projected onto higher $L_{\rm bol}$ values. In the luminosity range that we are considering, the AGN merger fraction is thus less affected by these uncertainties. To estimate the total XLF we take the sum of $\Phi_{\rm L, non-m}$ and $\Phi_{\rm L, m}$. The sum is dominated by non-merger galaxies whose space density is affected by significantly less uncertainty than the space density of major merger galaxies. For the majority of models and luminosities that we are considering, the uncertainty on the total XLF is thus smaller than the uncertainty on the AGN merger fraction.    

To compare our predictions to the observations we use the compilation of observations of the merger fraction by \citetalias{Treister:2012ab}. It combines merger fraction measurements for AGN that were selected in the optical \citep{Bahcall:1997aa, Koss:2010aa}, the X-rays \citep{Georgakakis:2009aa, Koss:2010aa, Cisternas:2011aa, Schawinski:2011ab, Kocevski:2012aa}, and in the infrared \citep{Urrutia:2008aa, Kartaltepe:2010aa, Schawinski:2012aa}. The \textit{Swift}/Burst Alert Telescope \citep{Gehrels:2004aa, Barthelmy:2005aa} X-ray luminosity function (XLF) by \citetalias{Ajello:2012aa} serves as an additional constraint. 

The galaxy major merger fraction is know to evolve with redshift (e.g. \citealt{Bridge:2010aa, Lotz:2011aa, Keenan:2014aa, Robotham:2014aa}). We thus concentrate on $0.02 < z < 0.06$, the redshift range that was used for the blue and green and major merger mass functions. \citetalias{Ajello:2012aa} used a sample of AGN with a median redshift of $0.029$ for their XLF. The \citetalias{Ajello:2012aa} XLF is thus suitable for a comparison of our predictions and the observations. Furthermore, the best-fitting X-ray ERDF by \citetalias{Weigel:2017aa}, which is based on the \citetalias{Ajello:2012aa} results, can be used for the $0.02 < z < 0.06$ range. \citetalias{Treister:2012ab} combine AGN merger fraction measurements over a wide redshift range ($0 < z < 3$). They could not establish a redshift dependence. We thus assume that the measurements by \citetalias{Treister:2012ab} are valid for the local Universe. 

\subsection{Aim}\label{sec:aim}
By varying the ERDF of AGN in major merger galaxies and by comparing our predictions to the observations, we are able to explore the luminosity dependence of the AGN merger fraction, the AGN host galaxy merger fraction and possible $\xi_{\rm m}(\lambda_{\rm Edd})$ shapes. Our aim is \textit{not} to find the merger ERDF that best reproduces the observations. It would be possible to use a maximum likelihood approach to determine the best-fitting $\xi_{\rm m}(\lambda_{\rm Edd})$ that allows us to reproduce both the observed AGN merger fraction and the total XLF. However, the large number of assumptions and the observational uncertainties do not justify such a sophisticated statistical analysis. For example, by combining multiple selection methods over a wide redshift range, \citetalias{Treister:2012ab} show the prominent luminosity dependence of the AGN merger fraction. On the one hand, the inhomogeneity of the sample implies that each measurement applies to a certain group of galaxies and is affected by different selection effects and uncertainties.  For instance, at low bolometric luminosities, the optical selection of AGN is impeded by star formation (e.g. \citealt{Jones:2016aa}). The \citetalias{Treister:2012ab} measurement at $\log (L_{\rm bol}/\rm erg\ s^{-1}) = 43.7$, which is based on the optical selection of AGN in the SDSS \citep{Koss:2010aa}, could thus be affected by incompleteness. In addition to the \textit{AGN} selection being susceptible to biases, the identification of merging \textit{galaxies} itself strongly depends on, for example, the depth of the images that are used \citep{van-Dokkum:2005aa, Schawinski:2010aa, Hong:2015aa}. On the other hand, the comparison of diverse selection methods, surveys and merger classifications, reveals the underlying luminosity dependence and provides valuable constraints for the AGN merger fraction measurement. We are thus not using a detailed quantitative analysis to determine the ERDF of AGN in major merger galaxies based on the \citetalias{Treister:2012ab} results. Rather, our aim is to show qualitatively what must broadly be true for $\xi_{\rm m}(\lambda_{\rm Edd})$. 
\subsection{Results}
\begin{figure*}
	\includegraphics[width=\textwidth]{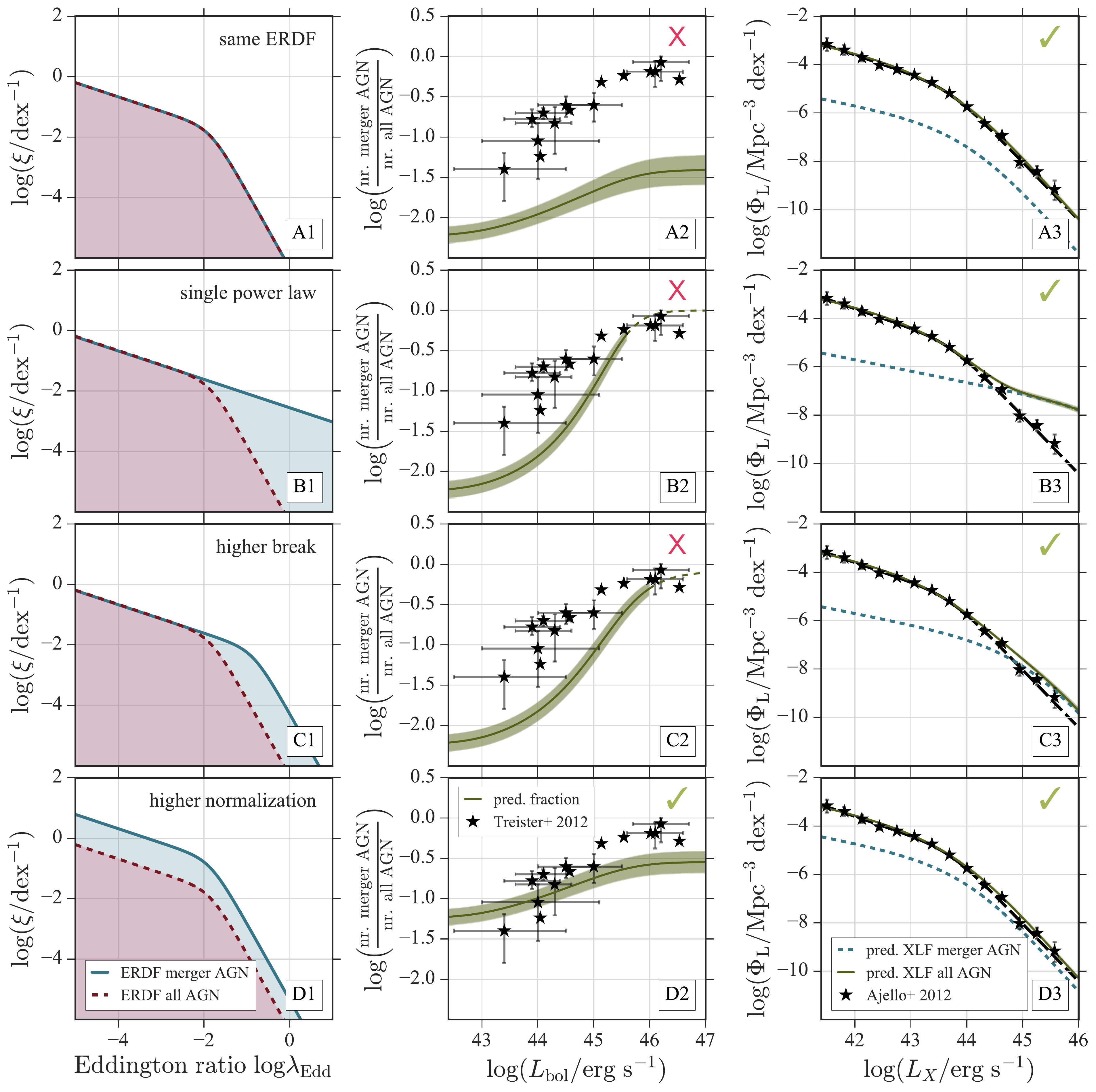}
	\caption{\label{fig:results1}Resulting AGN merger fractions and XLFs for four out of the six  different $\xi_{\rm m}(\lambda_{\rm Edd})$ shapes that we are considering. Each row represents one choice for the ERDF of AGN in major merger galaxies (blue solid line, left-hand side panels, also see Table \ref{tab:para}). The  ERDF of all radiatively efficient AGN (red dashed line, left-hand side panels) is kept constant. We predict the number of AGN in major merger galaxies relative to all AGN as a function of bolometric luminosity (central panels, green solid line, only computed where the predicted $\Phi_{\rm L, merg} < \Phi_{\rm L, all}$). We also show the observed AGN merger fraction (\protect\citetalias{Treister:2012ab}, black stars). In the right-hand side panels, we show our prediction for the total XLF (green solid lines, contribution of merger AGN to total XLF: blue dashed line). The observed XLF \protect\citepalias{Ajello:2012aa} is shown by the black solid line and the star-shaped markers. The green shaded areas show the uncertainty on the AGN merger fraction and the XLF. The green check marks represent the success, the red cross marks the failure a model to reproduce the observed AGN merger fraction and XLF (see text for details). The figure illustrates that the luminosity dependence of the AGN merger fraction arises, even if we assume the same ERDFs for merger AGN and for the general AGN population (model A). Moreover, a change in the normalization of $\xi_{\rm m}(\lambda_{\rm Edd})$ is necessary to match the observed AGN merger fraction (model D). Physically this can be interpreted as a higher AGN fraction among major merger galaxies. Increasing the fraction of high $\lambda_{\rm Edd}$ AGN, while keeping the integral over the merger ERDF fixed (models B and C), results in a AGN merger fraction that is inconsistent with the \protect\citetalias{Treister:2012ab} result and can lead to an overprediction of the number of bright AGN.}
\end{figure*}

\begin{figure*}
	\includegraphics[width=\textwidth]{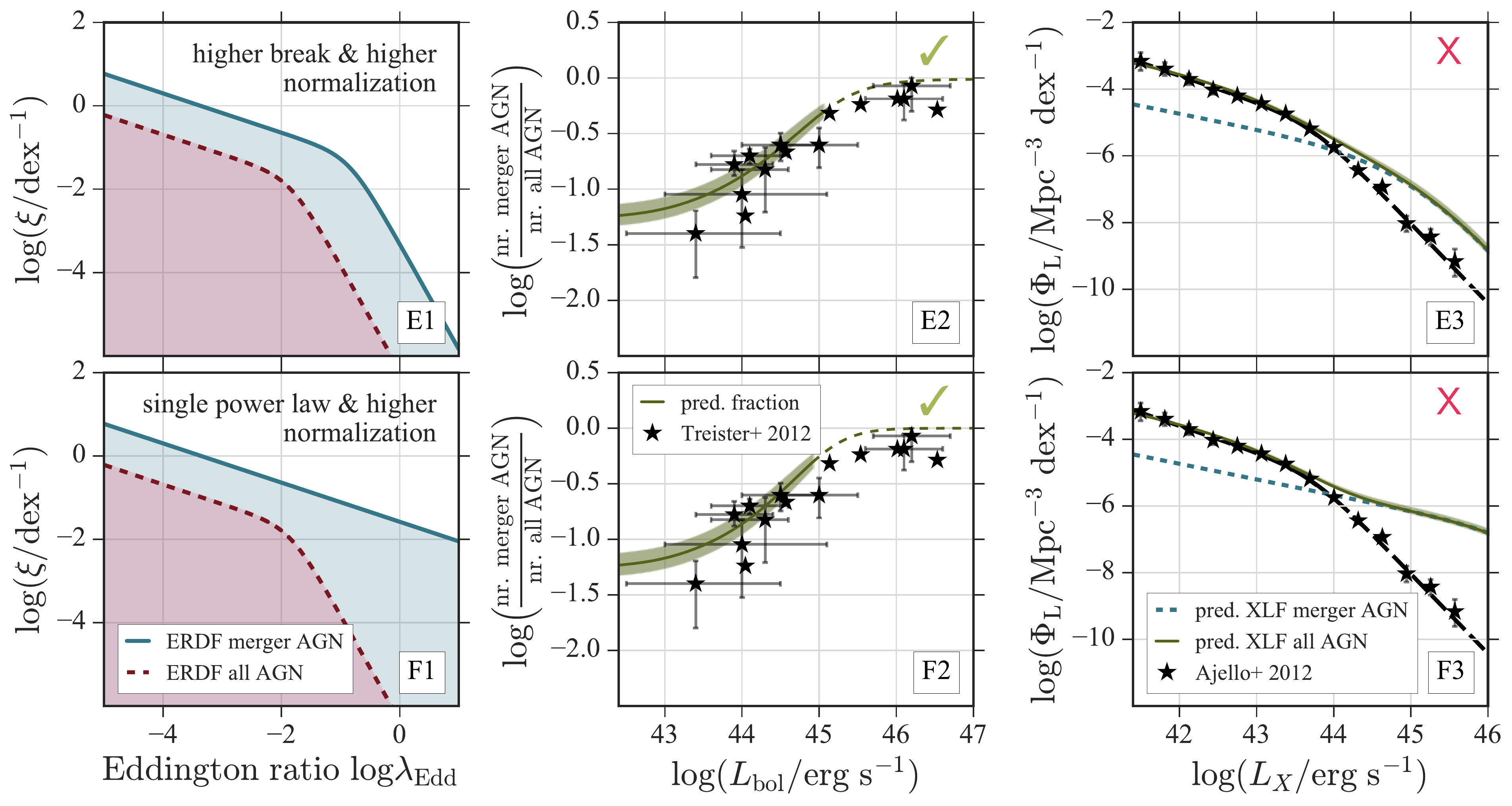}
	\caption{\label{fig:results2}Resulting AGN merger fractions and XLFs for the remaining two  $\xi_{\rm m}(\lambda_{\rm Edd})$ shapes that we are considering. Symbols are the same as for Fig. \ref{fig:results1}. For both $\xi_{\rm m}(\lambda_{\rm Edd})$ models shown here we have increased the normalization of the merger ERDF relative to the ERDF that we assume for all radiatively efficient AGN. The figure illustrates that an additional significant increase of the fraction of high $\lambda_{\rm Edd}$ AGN leads to an overprediction of bright AGN. While the predicted AGN merger fraction is consistent with the observations of \protect\citetalias{Treister:2012ab}, the predicted total XLF deviates significantly from the observations by \protect\citetalias{Ajello:2012aa}. $\lambda^{*}_{\rm m}$ and $\delta_{2, \rm m}$ values as extreme as the ones used here are thus unlikely.}
\end{figure*}

\setlength{\tabcolsep}{3pt}
\begin{table}
	\begin{tabular}{llllllll}
		\small
		{$\xi_{\rm all}(\lambda)$} & {$\log \lambda^{*}_{\rm Edd, all}$} & {$\delta_{1, \rm all}$} & {$\delta_{2, \rm all}$} & {} & {} & {} & {}\\
		\hline
		{} & {\lambdastarXnoerror} & {\deltaXanoerror} & {\deltaXbnoerror} & {} & {} & {} & {}\\
		{} & {} & {} & {} & {} & {} & {} & {}\\
		{$\xi_{\rm m}(\lambda)$} & {$\log \lambda^{*}_{\rm Edd, m}$} & {$\delta_{1, \rm m}$} & {$\delta_{2, \rm m}$} & {$\int \xi_{\rm m}$} & {$\xi^{*}_{\rm m}/\xi^{*}_{\rm all}$} & {$\chi^2_{\rm red, frac}$} & {$\chi^2_{\rm red, XLF}$}\\
		\hline
		{A} & {\lambdastarXnoerror} & {\deltaXanoerror} & {\deltaXbnoerror} & {$\int \xi_{\rm all}$} & {\modelAxistarratio} & {\modelAredchifrac} & {\modelAredchixlf}\\
		{B} & {\lambdastarXnoerror} & {\deltaXanoerror} & {\deltaXanoerror} & {$\int \xi_{\rm all}$} & {\modelBxistarratio} & {\modelBredchifrac} & {\modelBredchixlf}\\
		{C} & {\lambdastarXnoerror + 1} & {\deltaXanoerror} & {\deltaXbnoerror} & {$\int \xi_{\rm all}$} & {\modelCxistarratio} &  {\modelCredchifrac}& {\modelCredchixlf}\\
		{D} & {\lambdastarXnoerror} & {\deltaXanoerror} & {\deltaXbnoerror} & {$10 \times \int \xi_{\rm all}$} & {\modelDxistarratio} & {\modelDredchifrac} & {\modelDredchixlf}\\
		{E} & {\lambdastarXnoerror + 1} & {\deltaXanoerror} & {\deltaXbnoerror} & {$10 \times \int \xi_{\rm all}$} & {\modelExistarratio} & {\modelEredchifrac} & {\modelEredchixlf}\\
		{F} & {\lambdastarXnoerror} & {\deltaXanoerror} & {\deltaXanoerror} & {$10 \times \int \xi_{\rm all}$} & {\modelFxistarratio} & {\modelFredchifrac} & {\modelFredchixlf}\\
	 \end{tabular}
	 \caption{\label{tab:para}Parameter overview for the six merger ERDF shapes that we are considering. For the general galaxy population, we keep the shape of the ERDF, $\xi_{\rm all}(\lambda)$ constant and fixed to the parameters determined by \citetalias{Weigel:2017aa}. For AGN in major merger galaxies, we investigate the effects of six different $\xi_{\rm m}(\lambda)$ shapes. Please note that, our aim is not to find the best-fitting merger ERDF, but to probe the boundaries of parameter space and to show qualitatively what must and what cannot be true for $\xi_{\rm m}(\lambda)$. Compared to $\xi_{\rm all}(\lambda)$, we change the position of the break $\log \lambda^{*}$, the high Eddington ratio slope $\delta_2$ and the integral over $\xi_{\rm m}(\lambda)$. In the last two columns we summarize the reduced chi-squared values for our predictions which we compute relative to the observed AGN merger fraction by \citetalias{Treister:2012ab} and the total XLF by \citetalias{Ajello:2012aa}. As we discuss in the text, these $\chi^2_{\rm red}$ values should only be used for a general model comparison. }
\end{table}

We keep the ERDF of the general AGN population constant and consider six different shapes for the ERDF of AGN in major merger galaxies. These six cases represent what one might intuitively expect for the accretion properties of AGN in merger systems. In terms of ERDF parameters, these six scenarios probe the extreme boundaries of parameter space. By investigating their effect, we are able to exclude merger ERDF models, restrict parameter space, and show qualitatively what must and what cannot be true for the ERDF of AGN in major merger galaxies. 

In Fig. \ref{fig:results1} and \ref{fig:results2} we show the six different merger ERDFs and their effect on the AGN merger  fraction and the XLF. We refer to the six different scenarios as models A-F. The left-hand side columns show the assumed merger ERDFs in comparison to the X-ray ERDF by \citetalias{Weigel:2017aa}, which we use for the general AGN population. The central panels illustrate the resulting AGN merger fraction as a function of bolometric luminosity and the corresponding measurements by \citetalias{Treister:2012ab}. In the right-hand panels we show our prediction for the total XLF and the observed XLF by \citetalias{Ajello:2012aa}. 

We are not using a minimisation procedure to find the best-fitting merger ERDF model. To still be able to compare the goodness of fit of our different models, we calculate a simple reduced chi-squared ($\chi^2_{\rm red}$) for $\log (\rm AGN\ merger\ fraction)$ and $\log \Phi_{\rm L}$ for the XLF. We note that in Fig. \ref{fig:results1} and \ref{fig:results2} the AGN merger fraction and the XLF are shown on different scales. The uncertainties affecting the AGN merger fraction and the XLF thus differ less significantly than the figures may suggest. To include the errors on our predictions, we add the predicted and observed uncertainties in quadrature. When calculating the $\chi^2_{\rm red}$ value for the AGN merger fraction, we only consider luminosity bins where our predicted $\Phi_{\rm L, m} < \Phi_{\rm L, all}$ and neglect the observed uncertainty in terms of bolometric luminosity. Furthermore, the \citetalias{Treister:2012ab} compilation does not include uncertainties on the merger fraction measurements by \cite{Kartaltepe:2010aa}. For these data points thus only the predicted uncertainty contributes to $\chi^2_{\rm red}$. For the number of degrees of freedom we use $N_{\rm bins}$, the number of considered luminosity bins. Our model is non-linear which implies that the number of degrees of freedom is not well defined \citep{Andrae:2010aa}. Given this uncertainty and the fact that errors on the bolometric luminosity and the \cite{Kartaltepe:2010aa} measurements are being neglected, we stress that the computed $\chi^2_{\rm red}$ values should only be used for a general model comparison.

For each model and each $N_{\rm bins}$ value, we also determine the $3\sigma$ value of the corresponding chi-squared distribution. If the computed (not reduced) $\chi^2$ value lies within the $3\sigma$ boundary, we consider our prediction to be `a good fit'. In Fig. \ref{fig:results1} and \ref{fig:results2}, we mark these results with a green check mark. If the computed $\chi^2$ value lies beyond $3\sigma$, we flag the subplot with a red cross mark. 

For the following six models we consider different values for the break ($\lambda^{*}_{\rm Edd}$), the integral and thus also the normalization ($\xi^{*}_{\rm m}$), and the high Eddington ratio slope ($\delta_{2, \rm m}$) of the merger ERDF. The ERDF of the general AGN population remains fixed to the parameters determined by \citetalias{Weigel:2017aa}. We also do not change the low Eddington ratio slope of the merger ERDF. We set it to $\delta_{1, \rm m} =$ \deltaXanoerror, the $\delta_{1}$ value of the general AGN population. Note that, unless stated otherwise, $\xi_{\rm m}(\lambda_{\rm Edd})$ and $\xi_{\rm all}(\lambda_{\rm Edd})$ are normalized to have the same integral. According to equation \ref{eq:host_gal_merg_frac}, this implies the same AGN fraction for major merger galaxies and the general AGN population. Below, we will abbreviate the integral over the ERDF, i.e. $\int \xi (\lambda_{\rm Edd}) d\log \lambda_{\rm Edd}$,  as $\int \xi$. Table \ref{tab:para} summarizes the assumed ERDF parameters and resulting $\chi^2_{\rm red}$ values.
 
\begin{itemize}
	\item{\textbf{Model A}: same ERDF, $\xi_{\rm m}(\lambda_{\rm Edd}) = \xi_{\rm all}(\lambda_{\rm Edd})$: If we use the same ERDF for AGN in major mergers as for the general galaxy population, the predicted AGN merger fraction lies about one order of magnitude below the observed fraction ($\chi^2_{\rm red}$ = \modelAredchifrac). The predicted combined XLF is consistent with the observations ($\chi^2_{\rm red}$ = \modelAredchixlf).}
	
	\item{\textbf{Model B}: single power law, $\delta_{2, \rm m} = \delta_{1, \rm m}$: Using a single power law shaped merger ERDF results in too many bright AGN compared to the observations. The predicted and observed AGN merger fractions only agree at high luminosities ($\chi^2_{\rm red}$ = \modelBredchifrac). At the bright end, the predicted total XLF deviates from the observations ($\chi^2_{\rm red}$ = \modelBredchixlf). Nonetheless, the resulting $\chi^2$ value lies within $3\sigma$ of the corresponding chi-squared distribution. }
	
	\item{\textbf{Model C}: higher break, $\log \lambda_{\rm Edd, m}^{*} = \log \lambda_{\rm Edd, all}^{*} + 1$: By increasing the break of the merger ERDF by an order of magnitude, the predicted AGN merger fraction becomes steeper and is consistent with the observations at high luminosities ($\chi^2_{\rm red}$ = \modelCredchifrac). The predicted XLF is consistent with the observed XLF ($\chi^2_{\rm red}$ = \modelCredchixlf).}
	
	\item{\textbf{Model D}: higher normalization, $\int \xi_{\rm m} = 10 \times \int \xi_{\rm all}$: For this model we increase the integral over the merger ERDF by an order of magnitude compared to the integral over $\xi_{\rm all}(\lambda_{\rm Edd})$. As we have shown in equation \ref{eq:host_gal_merg_frac}, this implies a ten times higher AGN fraction among major merger galaxies than among the general galaxy population.  Such a $\xi_{\rm m}(\lambda)$ results in a AGN merger fraction that, within the probed luminosity range, is broadly consistent with the observations ($\chi^2_{\rm red}$ = \modelDredchifrac, $\chi^2$ within $3\sigma$). Due to the identical shape of $\xi_{\rm m}(\lambda)$ and $\xi_{\rm all}(\lambda)$, the total XLF is in agreement with the observations ($\chi^2_{\rm red} = $ \modelDredchixlf).}
	
	\item{\textbf{Model E}: higher break and higher normalization, $\log \lambda_{\rm Edd, m}^{*} = \log \lambda_{\rm  Edd, all}^{*} + 1$, $\int \xi_{\rm m} = 10 \times \int \xi_{\rm all}$: This ERDF merger model leads to a significant deviation from the observed XLF at the bright end ($\chi^2_{\rm red} =$\modelEredchixlf). Above $\log (L_{\rm X}/\rm erg\ s^{-1}) \sim 44.5$ the model predicts that the space density of merger AGN is higher than the space density of all AGN. At $\log (L_{\rm bol}/\rm erg\ s^{-1})\lesssim 45$ the predicted AGN fraction is in agreement with the observations ($\chi^2_{\rm red} = $ \modelEredchifrac).}
	
	\item{\textbf{Model F}: single power law and higher normalization, $\delta_{2, \rm m} = \delta_{1, \rm m}$ and $\int \xi_{\rm m} = 10 \times \int \xi_{\rm all}$: The results of this merger ERDF model are similar to the results of model E: the XLF significantly deviates from the observations at the bright end ($\chi^2_{\rm red} = $ \modelFredchixlf) and in the range where $\Phi_{\rm L, m} < \Phi_{\rm L, all}$ the AGN fraction is consistent with the observations ($\chi^2_{\rm red} =$ \modelFredchifrac).}
\end{itemize}

We conclude that the luminosity dependence of the AGN merger fraction arises even if we assume identical ERDFs for merger AGN and all AGN (see model A). The luminosity dependence of the AGN merger fraction is thus not purely caused by different ERDFs.  

To be consistent with the observations by \citetalias{Treister:2012ab}, the integral over the ERDF of major merger AGN must be higher than the integral over the ERDF of all radiatively efficient AGN. Only increasing the number of high accretion rate AGN among the merger galaxy population is insufficient (see models B and C). Model D shows that increasing the integral over $\xi_{\rm m}(\lambda_{\rm Edd})$ by a factor 10 relative to the integral over $\xi_{\rm all}(\lambda_{\rm Edd})$ leads to agreement between predictions and observations for both the AGN fraction and the XLF with the $\chi^2$ value lying within $3\sigma$ of the corresponding chi-squared distribution. Scaling factors of the order of $\sim 20$ result in a better fit to the observed AGN merger fraction at high $L_{\rm bol}$. Simultaneously, such normalization changes lead to an overestimation of the measurements at low bolometric luminosities, where the observations may be affected by AGN selection effects (see Sect. \ref{sec:aim}).

The fit can be improved (see $\chi^2_{\rm red}$ values in Table \ref{tab:para}) by increasing the position of the break or by decreasing the value of the high Eddington ratio slope in addition to the change in normalization. However, extreme changes of $\lambda^{*}_{\rm Edd, m}$ and $\delta_{2, \rm m}$  lead to a significant deviation from the observed XLF, as models E and F illustrate.

We conclude that a change in the normalization of $\xi_{\rm m}(\lambda)$ is necessary. An additional increase of $\log \lambda^{*}_{\rm Edd, m}$ or a shallower high Eddington ratio slope $\delta_{2, \rm m}$ is possible. However, $\lambda^{*}_{\rm Edd, m}$ and $\delta_{2, \rm m}$ values as extreme as the ones we tested in models E and F are unlikely. $\xi_{\rm m}(\lambda_{\rm Edd})$ is thus likely to have $\xi^{*}_{\rm m} > \xi^{*}_{\rm all}$ in combination with $\log \lambda^{*}_{\rm Edd, all} < \log \lambda^{*}_{\rm Edd, m} < \log \lambda^{*}_{\rm Edd, all} + 1$, or $\delta_{1, \rm all} < \delta_{2, \rm m}$, or both.

\section{Discussion}
\subsection{The luminosity dependence of the AGN merger fraction}
Our results show that the luminosity dependence of the AGN merger fraction arises, even if $\xi_{\rm m}(\lambda_{\rm Edd})$ and $\xi_{\rm all}(\lambda_{\rm Edd})$ are identical and merger AGN are not more likely to have high $\lambda_{\rm Edd}$ values. The luminosity dependence can be traced back to different $M^{*}$ values for major mergers and blue+green galaxies and a mass dependent galaxy major merger fraction \citepalias{Weigel:2017ab}. Compared to lower mass galaxies, more massive galaxies are more likely to experience a major galaxy merger. Having a higher stellar mass implies a higher black hole mass and for a given $\lambda_{\rm Edd}$ value, a higher bolometric luminosity. These massive galaxies hence dominate the measured AGN merger fraction at high luminosities. The \textit{mass} dependence of the \textit{galaxy merger fraction}, which according to \cite{Hopkins:2010aa,Hopkins:2010ab} might be due to the non-linear halo mass to stellar mass conversion, thus causes a  \textit{luminosity} dependence of the \textit{AGN merger fraction}. The observations by \citetalias{Treister:2012ab} and, for example, also \cite{Glikman:2015aa}, can thus, at least partially, be accounted for by host galaxy properties.

Our results imply a mass dependent AGN merger fraction. Following \citetalias{Weigel:2017aa}, we are using mass independent ERDFs. This implies that major merger galaxies on all mass scales are equally likely to host AGN. Similarly, the AGN fraction among the general galaxy population is mass independent. With increasing stellar mass, the fraction of galaxies that are found to be in major mergers increases \citepalias{Weigel:2017ab}. Thus, if (nr. mergers/nr. galaxies) is mass dependent, the number of AGN in merger galaxies relative to the number of AGN among the general galaxy population, that is the AGN merger fraction, has to increase as a function of stellar mass. This also implies that if the AGN merger fraction is purely driven by host galaxy properties and if we are only considering \textit{one stellar mass bin}, the AGN merger fraction is expected to be luminosity \textit{independent}.
 
\subsection{The AGN fraction of merger galaxies vs. the AGN fraction of the general galaxy population}
Our results show that to match the observed AGN merger fraction, the integral of the merger ERDF has to be increased by a factor of $\sim10$ relative to the ERDF of all radiatively efficient AGN. This significant increase in $\xi^{*}_{\rm m}$ is necessary as there are significantly fewer major merger than blue+green galaxies\footnote{$\log(\Phi^{*}_{\rm m}/\Phi^{*}_{\rm bg}) \sim -1.85$, whereas $\log (\Phi^{*}_{\rm L, m}/\Phi^{*}_{\rm L, all}) \sim -0.75$ according to \citetalias{Treister:2012ab}}. Physically this can be interpreted as a higher AGN fraction among major merger galaxies: compared to the general galaxy population, major merger galaxies are $\sim 10$ times more likely to host AGN (see equation \ref{eq:host_gal_merg_frac}). 

Similar results have been found observationally by \cite{Goulding:2017aa}. Using a sample of infrared selected AGN merger systems in the range $0.1 < z < 0.9$, they show that independent of stellar mass and redshift, major merger galaxies are $\sim 2-7$ times more likely to host AGN than non-merger systems. Contradictory to our expectation, they find this value to be mass independent. Their results also imply that the majority of $\log (L_{\rm bol}/\rm erg\ s^{-1}) > 45$ AGN are likely to be found in major merger galaxies. At higher redshift, \cite{Hewlett:2017aa} draw contradictory conclusions. Based on a sample of X-ray selected AGN in the redshift range $0.5 < z < 2.2$, they find the fraction of AGN among major merger galaxies to neither be enhanced relative non-merger systems, nor to be luminosity dependent. 

We note that, our method is not based on measuring the AGN fraction among merger and non-merger galaxies. In contrast to classical observational approaches, our forward modeling analysis does not require the construction of a comparison sample of non-merger galaxies. Instead, the observed luminosity dependence of the AGN merger fraction is a direct consequence of the input stellar mass functions. 

\citetalias{Weigel:2017aa} argue that the shape of the ERDF might be determined by the efficiency with which gas is driven from the inner few parsecs to the accretion disc, the properties of this gas or both. We conclude that $\xi_{\rm m}(\lambda_{\rm Edd})$ is likely to have a significantly higher $\xi^{*}_{\rm m}$, but might resemble $\xi_{\rm all}(\lambda)$ in terms of its shape. In this context, our results   could imply that for merger AGN the physical conditions within the inner few parsecs are similar to those of AGN that are triggered by secular processes. However, due to gas being driven to the nuclei of galaxies (i.e. $\sim 100$ pc) and more fuel being available, black holes in major merger galaxies are more likely to `switch on' and enter a AGN phase. This has previously been suggested by, for instance, \cite{Sabater:2015aa}. Furthermore, the results by, for example, \cite{Koss:2012aa} and \cite{Fu:2018aa} imply that, if the interacting galaxies are gas-rich, mergers are likely to even promote dual AGN activity.

\subsection{The ERDF of AGN in major merger galaxies}
Fig. \ref{fig:results1} and \ref{fig:results2} show that a change in $\lambda_{\rm Edd, m}^{*}$ and $\delta_{2, \rm m}$ in addition to the increase of $\int \xi_{\rm m}$ relative to the integral over the ERDF for the general AGN is possible. Compared to the ERDF for the general AGN population, the ERDF of AGN in major merger galaxies could thus have a higher break $\lambda^{*}_{\rm Edd, m}$ or a shallower slope $\delta_{2, \rm m}$. However, an increase as extreme as what we have assumed for models E and F leads to inconsistency with the observed XLF and is thus unlikely. This result is non-trivial, as, even after increasing the AGN fraction among merger galaxies, one might have expected the space density of merger AGN to be too low to significantly affect the total XLF. While we did not use a minimisation approach to find the best-fitting merger ERDF, or qualitative approach did thus allow us to probe and constrain the extreme boundaries of parameter space.

As we discussed above, the necessary increase regarding the integral over the merger ERDF implies a $\sim 10$ times higher AGN fraction among major merger galaxies than among the general galaxy population. Compared to all AGN, the probability of hosting an AGN and accreting at any given $\lambda_{\rm Edd}$ is thus higher for AGN in major merger galaxies. However, since the merger ERDF is likely to resemble $\xi_{\rm all}(\lambda_{\rm Edd})$ in terms of shape, AGN in merger galaxies are unlikely to accrete at significantly higher Eddington ratios than AGN among the general galaxy population. 

This conclusion is based on the assumption that we are capable of detecting all merger AGN in the hard X-rays. If AGN in major merger galaxies are heavily Compton-thick, they could be missed by the hard X-ray selection and could be excluded from the XLF measurement. This could resolve the inconsistency between our predictions and the observed XLF for models E and F. The hard X-ray selection might indeed be incomplete regarding Compton-thick AGN \citep{Ricci:2015aa} and a fraction of merger AGN might experience a phase of Compton-thick black hole growth \citep{Kocevski:2015aa}. However, as there is no evidence for the Compton-thick fraction being luminosity dependent \citep{Ricci:2017ab}, we would expect to underestimate the space density of X-ray AGN at all luminosities, not just at the bright end of the XLF. Moreover, if the true XLF is indeed as high as predicted by models E and F, then this may challenge the ability to account for the integrated mass density of SMBHs (i.e., the so-called \citeauthor{Soltan:1982aa} argument; also see \citealt{Marconi:2004aa},  \citealt{Shankar:2009aa}).

We also assume that $\xi_{\rm m}(\lambda_{\rm Edd})$ and $\xi_{\rm all}(\lambda_{\rm Edd})$ have broken power law shapes. Instead of having two broken power law shaped ERDFs, one could imagine that AGN triggered through major mergers and secular processes dominate at high and low $\lambda_{\rm Edd}$ values, respectively. This is mathematically possible. However, to reproduce the observed XLF, the two ERDFs would have to be complementary and in particular, would have to have similar $\lambda_{\rm Edd}^{*}$ values. Furthermore, a $\xi_{\rm m}(\lambda_{\rm Edd})$ with a negative $\delta_1$ slope that drops at $\lambda_{\rm Edd} < \lambda_{\rm Edd}^{*}$ leads to a shallow faint end for the bolometric luminosity function of AGN in mergers (\citetalias{Weigel:2017aa}). This results in a steeply rising AGN merger fraction, incompatible with the \citetalias{Treister:2012ab} results. The observations thus provide tight constraints and using ERDFs that have the same functional form represents the simplest assumption possible. 

In the context of AGN downsizing \citep{Cowie:1996aa, Ueda:2003aa, Hasinger:2005aa} the redshift evolution of the luminosity dependence of the AGN merger fraction is of particular interest. For the results by \citetalias{Treister:2012ab} a potential $z$ dependence could not be established. To keep $\Phi_{\rm L, m}(\log L_{\rm bol})/\Phi_{\rm L, all}(\log L_{\rm bol})$ constant, the increase in the fraction of galaxies experiencing a merger with $z$ (e.g. \citealt{Bridge:2010aa, Lotz:2011aa, Keenan:2014aa, Robotham:2014aa}) and the resulting higher number of merger AGN has to be balanced by a higher AGN fraction among the general galaxy population \citep{Aird:2017aa}. This would imply a redshift dependent $\xi^{*}_{\rm m}/\xi^{*}_{\rm all}$ ratio, possibly contradicting a physical origin for the difference of the ERDF normalizations. 

To test a possible redshift dependence of $\xi_{\rm m}(\lambda_{\rm Edd})$ and $\xi_{\rm all}(\lambda_{\rm Edd})$ with our model, we would need well constrained, redshift resolved, merger mass functions. The \textit{integrated} galaxy merger fraction has been measured over a range of redshifts (e.g. \citealt{Bridge:2010aa, Lotz:2011aa, Keenan:2014aa, Robotham:2014aa}). However, measuring the mass function of major galaxy mergers requires at least of the order of 300 (\citetalias{Weigel:2017ab}) major merger galaxies with available stellar mass measurements. Our analysis would also greatly benefit from smaller uncertainties on the AGN merger fraction as a function of AGN luminosity and a uniform selection of mergers, AGN, and AGN in mergers over a wide luminosity range, at any given redshift.

As an alternative to our approach, $\xi_{\rm m}(\lambda_{\rm Edd})$ and $\xi_{\rm all}(\lambda_{\rm Edd})$ can be constrained observationally. Previous studies have constrained $\xi_{\rm all}(\lambda_{\rm Edd})$ for luminous, optically selected quasars in the local Universe and at higher $z$ (e.g. \citealt{Greene:2007aa,Schulze:2010aa,Kelly:2013aa,Schulze:2015aa}). However, a direct measurement of $\xi_{\rm m}(\lambda_{\rm Edd})$ is challenging as it requires hundreds of AGN in merger galaxies with $M_{\rm BH}$ and $\lambda_{\rm Edd}$ measurements.

The \textit{Swift}-BAT AGN Spectroscopic Survey (BASS, \citealt{Koss:2017aa}) provides optical spectra for several hundreds (>600) of hard X-ray selected AGN at $z\lesssim0.2$. With a well understood, unbiased AGN selection, this sample will allow us to put new constraints on the luminosity dependence of the AGN merger fraction. The $M_{\rm BH}$ and $\lambda_{\rm Edd}$ measurements provided within BASS will enable us to directly constrain $\xi_{\rm all}(\lambda_{\rm Edd})$ for local AGN, regardless of their level of obscuration. Even though the number statistics are still too small to allow a direct measurement of $\xi_{\rm m}(\lambda_{\rm Edd})$, BASS will allow us to gain a more complete census of the local AGN population and a better understanding of the AGN-merger connection.

\section{Summary}
We use the stellar mass function of major merger galaxies \citepalias{Weigel:2017aa} and a phenomenological model which links the galaxy to the AGN population \citepalias{Weigel:2017ab} to investigate the correlation between the occurrence of major galaxy mergers and the triggering of AGN in the local Universe. We predict the AGN merger fraction, that is the fraction of AGN in merger galaxies relative to all AGN, and the AGN X-ray luminosity function (XLF) of all AGN. We compare our predictions to the observations by \citetalias{Treister:2012ab} and \citetalias{Ajello:2012aa}. Our qualitative analysis shows that:

\begin{enumerate}
	\item \textbf{The luminosity dependence of the AGN merger fraction arises, at least partially, due to host galaxy properties.} More massive galaxies being more likely to undergo a major merger \citepalias{Weigel:2017ab}, causes the AGN merger fraction to increase as a function of bolometric luminosity. This is true even if we assume the same Eddington ratio distribution function (ERDF) for merger AGN and all AGN (see model A, Fig. \ref{fig:results1}). 
	
	\item \textbf{The AGN fraction among major merger galaxies is likely to be higher than among the general galaxy population.} In our model this is evident since the integral over the ERDF of merger AGN has to be increase by a factor of $\sim 10$ relative to the integral over the ERDF of all AGN to match the observations by \citetalias{Treister:2012ab}  (see model D, Fig. \ref{fig:results1}). Fixing the integral over the ERDF of merger AGN, but increasing the fraction of high Eddington ratio AGN, leads to inconsistency with the \citetalias{Treister:2012ab} results (see models B and C, Fig. \ref{fig:results1}). 
	
	\item \textbf{Even though major merger galaxies are more likely to host AGN in general, these AGN are unlikely to accrete at significantly higher Eddington ratios than AGN among the general galaxy population.}  In our model, increasing the fraction of high Eddington ratio AGN in addition to increasing the integral over the ERDF is possible. However, changes as extreme as the ones assumed for models E and F (see Fig. \ref{fig:results2}), are unlikely as they overpredict the space density of  luminous AGN and thus lead to inconsistency with the observed XLF. The ERDF of AGN in major merger galaxies is thus likely to have a higher normalization, but  resemble the ERDF of all AGN in terms of shape.
\end{enumerate}

\section*{Acknowledgements}
AKW and KS gratefully acknowledge support from Swiss National Science Foundation Grants PP00P2\_138979 and PP00P2\_166159 and the ETH Zurich Department of Physics. ET acknowledges support from CONICYT grant Basal-CATA PFB-06/2007 and FONDECYT Regular 1160999. This research made use of NASA's ADS Service. This publication made use of Astropy, a community-developed core \textsc{Python} package for Astronomy \citep{2013A&A...558A..33A} and the \textsc{Python} plotting package matplotlib \citep{Hunter:2007}.

\bibliographystyle{mnras}
\bibliography{Bib_Lib.bib}


\bsp	
\label{lastpage}
\end{document}